\documentclass[sigconf]{acmart}

\usepackage{booktabs} 

\begin{document}
\title{Determining the Scale of Impact from Denial-of-Service Attacks in Real Time Using Twitter}

\copyrightyear{2018} 
\acmYear{2018} 
\setcopyright{acmcopyright}
\acmConference[DYNAMICS '18]{DYnamic and Novel Advances in Machine Learning and Intelligent Cyber Security Workshop}{December 3--4, 2018}{San Juan, PR, USA}
\acmBooktitle{DYnamic and Novel Advances in Machine Learning and Intelligent Cyber Security Workshop (DYNAMICS '18), December 3--4, 2018, San Juan, PR, USA}
\acmPrice{15.00}
\acmDOI{10.1145/3306195.3306199}
\acmISBN{978-1-4503-6218-4/18/12}

\author{Chi Zhang}

\affiliation{%
  \institution{University of Maryland, Baltimore County}
  \department{Department of Computer Science and Electrical Engineering}
  \streetaddress{1000 Hilltop Circle}
  \city{Baltimore}
  \state{MD}
  \postcode{21250}
}
\email{chzhang1@umbc.edu}

\author{Bryan Wilkinson}

\affiliation{%
  \institution{University of Maryland, Baltimore County}
  \department{Department of Computer Science and Electrical Engineering}
  \streetaddress{1000 Hilltop Circle}
  \city{Baltimore}
  \state{MD}
  \postcode{21250}
}
\email{bwilk1@umbc.edu}
\author{Ashwinkumar Ganesan}

\affiliation{%
  \institution{University of Maryland, Baltimore County}
  \department{Department of Computer Science and Electrical Engineering}
  \streetaddress{1000 Hilltop Circle}
  \city{Baltimore}
  \state{MD}
  \postcode{21250}
}
\email{gashwin1@umbc.edu}
\author{Tim Oates}

\affiliation{%
  \institution{University of Maryland, Baltimore County}
  \department{Department of Computer Science and Electrical Engineering}
  \streetaddress{1000 Hilltop Circle}
  \city{Baltimore}
  \state{MD}
  \postcode{21250}
}
\email{oates@umbc.edu}

\begin{abstract}
Denial of Service (DoS) attacks are common in on-line and mobile services such as Twitter, Facebook and banking. As the scale and frequency of Distributed Denial of Service (DDoS) attacks increase, there is an urgent need for determining the impact of the attack. Two central challenges of the task are to get feedback from a large number of users and to get it in a timely manner. In this paper, we present a weakly-supervised model that does not need annotated data to measure the impact of DoS issues by applying Latent Dirichlet Allocation and symmetric Kullback-Leibler divergence on tweets. There is a limitation to the weakly-supervised module. It assumes that the event detected in a time window is a DoS attack event. This will become less of a problem, when more non-attack events twitter got collected and become less likely to be identified as a new event. Another way to remove that limitation, an optional classification layer, trained on manually annotated DoS attack tweets, to filter out non-attack tweets can be used to increase precision at the expense of recall. Experimental results show that we can learn weakly-supervised models that can achieve comparable precision to supervised ones and can be generalized across entities in the same industry.
\end{abstract}



\maketitle

\section{Introduction}

Denial of Service attacks are explicit attempts to stop legitimate users from accessing specific network systems \cite{carl2006denial}. Attackers  try to exhaust network resources like bandwidth, or server resources like CPU and memory. As a result, the targeted system slows down or becomes unusable \cite{mirkovic2004taxonomy}. On-line service providers like Bank Of America, Facebook and Reddit are often the target of such attacks and the frequency and scale of those attacks has increased rapidly in recent years \cite{zargar2013survey}.

To address this problem, there is ample previous work on methods to detect and handle Denial of Service attacks, especially Distributed Denial of Service attacks. D-WARD \cite{mirkovic2003source} is a scheme that tries to locate a DDoS attacks at the source by monitoring inbound and outbound traffic of a network and comparing it with predefined "normal" values. Some IP Traceback mechanisms \cite{john2009ddos} were developed to trace back to the attack source from the victim's end. Still other methods try to deploy a defensive scheme in an entire network to detect and respond to an attack at intermediate sub-networks. Watchers \cite{bradley1998detecting} is an example of this approach.

Despite all the new models and techniques to prevent or handle cyber attacks, DDoS attacks keep evolving. Services are still being attacked frequently and brought down from time to time. After a service is disrupted, it is crucial for the provider to assess the scale of the outage impact.


In this paper, we present a novel approach to solve this problem. No matter how complex the network becomes or what methods the attackers use, a denial of service attack always results in legitimate users being unable to access the network system or slowing down their access and they are usually willing to reveal this information on social media plaforms. Thus legitimate user feedback can be a reliable indicator about the severity level of the service outage. Thus we split this problem into two parts namely by first isolating the tweet stream that is likely related to a DoS attack and then measuring the impact of attack by analyzing the extracted tweets.

A central challenge to measure the impact is how to figure out the scale of the effect on users as soon as possible so that appropriate action can be taken. Another difficulty is given the huge number of users of a service, how to effectively get and process the user feedback. With the development of Social Networks, especially micro blogs like Twitter, users post many life events in real time which can help with generating a fast response. Another advantage of social networks is that they are widely used. Twitter claims that they had 313 million monthly active users in the second quarter of 2016 \cite{Twitter_Num_of_user}. This characteristic will enlarge the scope of detection and is extremely helpful when dealing with cross domain attacks because tweets from multiple places can be leveraged. The large number of users of social networks will also guarantee the sensitivity of the model. However, because of the large number of users, a huge quantity of tweets will be generated in a short time, making it difficult to manually annotate the tweets, which makes unsupervised or weakly-supervised models much more desirable. 

In the Twitter data that we collected there are three kinds of tweets. Firstly are tweets that are actually about a cyberattack. For example, someone tweeted "Can't sign into my account for bank of America after hackers infiltrated some accounts."  on September 19, 2012 when a attack on the website happened. Secondly are tweets about some random complaints about an entity like "Death to Bank of America!!!! RIP my Hello Kitty card... " which also appeared on that day. Lastly are tweets about other things related to the bank.  For example, another tweet on the same day is "Should iget an account with bank of america or welsfargo?".

To find out the scale of impact from an attack, we must first pick out the tweets that are about the attack. Then using the ratio and number of attack tweets, an estimation of severity can be generated. To solve the problem of detecting Denial of Service attacks from tweets, we constructed a weakly-supervised Natural Language Processing (NLP) based model to process the feeds. More generally, this is a new event detection model. We hypothesize that new topics are attack topics. The hypothesis would not always hold and this issue will be handled by a later module. The first step of the model is to detect topics in one time window of the tweets using Latent Dirichlet Allocation \cite{blei2003latent}. Then, in order to get a score for each of the topics, the topics in the current time window are compared with the topics in the previous time window using Symmetric Kullback-Leibler Divergence (KL Divergence) \cite{kullback1997information}. After that, a score for each tweet in the time window is computed using the distribution of topics for the tweet and the score of the topics. We're looking for tweets on new topics through time.  While the experiments show promising results, precision can be further increased by adding a layer of a supervised classifier trained with attack data at the expense of recall.

Following are the contributions in this paper:
\begin{enumerate}
\item A dataset of annotated tweets extracted from Twitter during DoS attacks on a variety organizations from differing domains such as banking (like \textit{Bank Of America}) and technology.
\item A weakly-supervised approach to identifying detect likely DoS service related events on twitter in real-time.
\item A score to measure impact of the DoS attack based on the frequency of user complaints about the event.
\end{enumerate}

The rest of this paper is organized as follows: In section 2, previous work regarding DDoS attack detection and new event detection will be discussed. In section 3, we describe the how the data was collected. We also present the model we created to estimate the impact of DDoS attacks from Twitter feeds. In section 4, the experiments are described and the results are provided. In section 5 we discuss some additional questions. Finally, section 6 concludes our paper and describes future work.

\section{Related Work}
Denial of Service (DoS) attacks are a major threat to  Internet security, and detecting them has been a core task of the security community for more than a decade. There exists significant amount of prior work in this domain. \cite{feinstein2003statistical,lee2008ddos,braga2010lightweight} all introduced different methods to tackle this problem. The major difference between this work and previous ones are that instead of working on the data of the network itself, we use the reactions of users on social networks to identify an intrusion.

Due to the widespread use of social networks, they have become an important platform for real-world event detection in recent years \cite{Goswami2016}. \citet{dou2012event} defined the task of new event detection as "identifying the first story on topics of interest through constantly monitoring news streams". Atefeh et al. \cite{atefeh2015survey} provided a comprehensive overview of event detection methods that have been applied to twitter data. We will discuss some of the approaches that are closely related to our work. Weng et al. \cite{weng2011event} used a wavelet-signal clustering method to build a signal for individual words in the tweets that was dependent high frequency words that repeated themselves. The signals were clustered to detect events. Sankaranarayanan et al. \cite{sankaranarayanan2009twitterstand} presented an unsupervised news detection method based on naive Bayes classifiers and on-line clustering. \citet{long2011towards} described an unsupervised method to detect general new event detection using Hierarchical divisive clustering. Phuvipadawat et al. \cite{phuvipadawat2010breaking} discussed a pipeline to collect, cluster, rank tweets and ultimately track events. They computed the similarity between tweets using TF-IDF. The Stanford Named Entity Recognizer was used to identify nouns in the tweets providing additional features while computing the TF-IDF score. Petrovi{\'c} et al. \cite{petrovic2010streaming} tried to detect events on a large web corpus by applying a modified locality sensitive hashing technique and clustering documents (tweets) together. Benson et al. \cite{benson2011event} created a graphical model that learned a latent representation for twitter messages, ultimately generating a canonical value for each event. Tweet-scan \cite{capdevila2017tweet} was a method to detect events in a specific geo-location. After extracting features such as name, time and location from the tweet, the method used DB-SCAN to cluster the tweets and Hierarchical Dirichlet Process to model the topics in the tweets. Badjatiya et. al. \cite{badjatiya2017deep} applied deep neural networks to detect events. They showed different architectures such as Convolutional Neural Networks (CNN), Recurrent Neural Networks (LSTM based) and FastText outperform standard n-gram and TF-IDF models. Burel et al. \cite{burel2017semantics} created a Dual-CNN that had an additional channel to model the named entities in tweets apart from the pretrained word vectors from GloVe \cite{pennington2014glove} or Word2Vec \cite{mikolov2013efficient}. 

Thus most event detection models can be grouped into three main categories of methods i.e. TF-IDF based methods, approaches that model topics in tweets and deep neural network based algorithms. One of the main challenges against applying a neural network model is the the requirement of a large annotated corpus of tweets. Our corpus of tweets is comparatively small. Hence we build our pipeline by modeling the topics learned from tweets.

The previous work that is most similar to ours was \citet{cordeiro2012twitter}. We both used Latent Dirichlet Allocation (LDA) to get the topics of the document, the difference was they only run LDA on the hash-tag of the tweets while we try to get the topics in the tweets by running it on the whole document. 

Latent Dirichlet Allocation \cite{blei2003latent} was a method to get topics from a corpus. In our work, we used the technique to acquire the values of some of the variables in our equation. A variation of it, Hierarchically Supervised Latent Dirichlet Allocation \cite{perotte2011hierarchically} was used in the evaluation.


\section{Approach}

\begin{figure*}
\includegraphics[width=\textwidth]{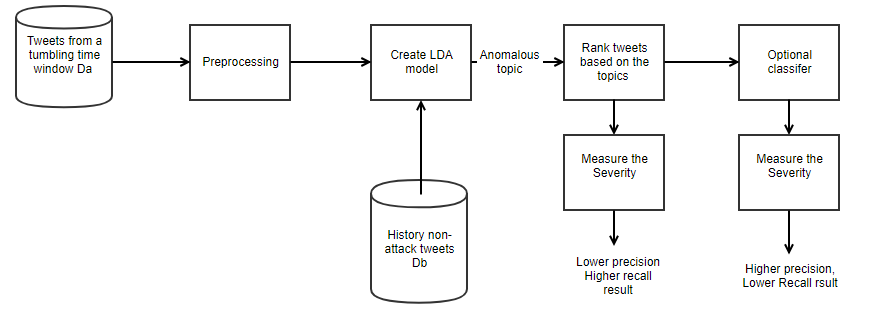}
\caption{Workflow to process tweets gathered and build a model to rank future tweets that likely to be related to a DoS attack. The ranked tweets are used to measure the severity of the attack.}
\centering
\label{flowchart}
\end{figure*}

Figure \ref{flowchart} outlines the entire pipeline of the model from preprocessing tweets to modeling them and finally detecting / ranking future tweets that are related to a DoS issue and measuring its severity.

\subsection{Data Collection}

To collect the tweets, we first gathered a list of big DDoS attacks happened from 2012 to 2014. Then for each attack on the list, we collected all the tweets from one week before the attack to the attack day that contains the name of the entity attacked.

\subsection{Preprocessing}
The following preprocessing procedure were applied to the corpus of tweets:
\begin{itemize}
\item Remove all the meta-data like time stamp, author, and so on. These meta-data could provide useful information, but only the content of the tweet was used for now.
\item Lowercase all the text
\item Use an English stop word list to filter out stop words. 
\end{itemize}
The last two steps are commonly used technique when preprocessing text.

\subsection{Create LDA Models}
Now we try to find out a quantitative representation of the corpus. To do that, the preprocessed tweets about one attack will be divided into two groups. One is on the attack day and the other is the tweets one week before it. The first set will be called $D_a$ and the other one $D_b$. This step will create two separate LDA models for $D_a$ and $D_b$ using the Genism library \cite{rehurek_lrec}. The first Model will be called $M_a$ and the other one $M_b$.

\begin{figure}[h!]
\includegraphics[width=0.5\textwidth]{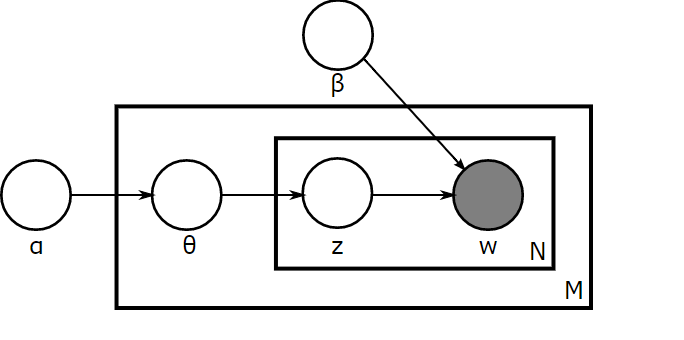}
\caption{Plate notation of LDA \protect\cite{blei2003latent}. The outer box denotes documents in the corpus and $M$ is the number of documents. The inner box denotes the repeated choice of topics and words within a document where $N$ is the number of words in a document. $\alpha$ is the parameter of the Dirichlet prior on the per-document topic distributions. $\beta$ is the parameter of the Dirichlet prior on the per-topic word distribution. $\theta$ is the topic distribution. $z$ is the topic of word $w$ in the document.}
\centering
\label{LDA}
\end{figure}

Latent Dirichlet allocation (LDA) is a generative probabilistic topic modeling model. Figure \ref{LDA} is its plate notation. The meaning of different parameters $M$, $N$, $\alpha$, $\beta$, $\theta$, $z$ and $w$ is also described there.

We used the LDA algorithm implemented by the Gensim library. One of the most important parameters of the LDA algorithm is the number of topics $N_t$ in the corpus. To determine that we introduced the following formula:
\begin{equation}
N_t= \left \lfloor{\alpha*logN_d}\right \rfloor
\end{equation}
where $N_d$ is the number of tweets in the corpus. $\alpha$ is a constant and we used $\alpha$=10 in our experiments. The logic behind the equation is discussed in section 5.

\subsection{The attack topics}

Then we want to find out how the new topics are different from the history topics or, in other words, how topics in $M_a$ differ from topics in $M_b$. We define the Symmetric Kullback-Leibler divergence for topic $T_j$ in Model $M_a$ as:

\begin{equation}
SKL_j=\min_{1<m<n} (D_{kl} (T_j,T_m^{'}) + D_{kl} (T_m^{'}, T_j ))
\end{equation}
Where n is the number of topics in Model $M_b$,  $T_m^{'}$ is the $m^{th}$ topic in Model $M_b$ and $D_kl (X,Y)$ is the original Kullback-Leibler Divergence for discrete probability distributions which defined as :
\begin{equation}
D_{kl} (X,Y)=\sum_i X(i)*\log {X(i)\over Y(i)}
\end{equation}
Where $X(i)$ and $Y(i)$ are the probability of token $i$ in topics $X$ and $Y$ respectively. This is similar to the Jensen-Shannon divergence.

So for each topic $T_j$ in Model $M_a$ its difference to topics in $M_b$ is determined by its most similar topic in $M_b$.

The topics from the attack day model $M_a$ are ranked by their Symmetric Kullback-Leibler divergence to topics from the non-attack day model $M_b$. An example of selected attack topics is provided in section 4.3.

\subsection{The attack tweets}

This subsection is about how to find specific tweets that are about a network attack. The tweets are selected based on the relative score $S$. The score for tweet $t_i$ is defined as:
\begin{equation}
S= \sum_{j=1}^n P_{i,j} * SKL_j 
\end{equation}
Where $n$ is the number of topics on the attack day, $P_{i,j}$ is the probability that topic $j$ appears in tweet $t_i$ in the attack day LDA model, and $SKL_j$ is the Symmetric Kullback-Leibler divergence for topic $j$. The higher the score the more likely it is related to an attack event.

\subsection{Optional Classifier Layer}
Because annotated data is not needed, the model we described before can be regarded as a weakly-supervised model to detect new events on twitter in a given time period. To label tweets as attack tweets, one assumption must be true, which is that the new event in that time period is a cyber attack. Unfortunately, that is usually not true. Thus, an optional classifier layer can be used to prevent false positives.

By using a decision tree model we want to find out whether the weakly-supervised part of the model can simplify the problem enough that a simple classification algorithm like a decision tree can have a good result. Additionally, it is easy to find out the reasoning underline a decision tree model so that we will know what the most important features are.

The decision tree classifier is trained on the bag of words of collected tweets and the labels are manually annotated. We limit the minimum samples in each leaf to be no less than 4 so that the tree won't overfit. Other than that, a standard Classification and Regression Tree (CART) \cite{breiman1984classification} implemented by scikit-learn \cite{scikit-learn} was used. The classifier was only trained on the training set (tweets about Bank of America on 09/19/2012), so that the test results do not overestimate accuracy.

\subsection{Measure the Severity}

The definition of severity varies from different network services and should be studied case by case.

For the sake of completeness, we propose this general formula:
\begin{equation}
SeverityLevel = \beta * {N_{attack} \over N_{all}} + (1 - \beta) * {N_{attack} \over N_{user}}
\end{equation}

In the equation above, $\beta$ is a parameter from 0 to 1 which determines the weight of the two parts. $N_{attack}$ is the number of attack tweets found. $N_{all}$ means the number of all tweets collected in the time period. And $N_{user}$ is the number of twitter followers of the network service.  

An interesting future work is to find out the quantitative relation between SeverityLevel score and the size of the actual DDoS attack.

\section{Experiments}
In this section we experimentally study the proposed attack tweet detection models and report the evaluation results.

\subsection{Term Definition}

We used precision and recall for evaluation:

\begin{itemize}
\item Precision: Out of all of the tweets that are marked as attack tweets, the percentage of tweets that are actually attack tweets. Or true positive over true positive plus false positive.
\item Recall: Out of all of the actual attack tweets, the percentage of tweets that are labeled as attack tweets. Or true positive over true positive plus false negative.
\end{itemize}

\subsection{Experiment Dataset}

We collected tweets related to five different DDoS attacks on three different American banks. For each attack,  all the tweets containing the bank's name posted from one week before the attack until the attack day were collected. There are in total 35214 tweets in the dataset. Then the collected tweets were preprocessed as mentioned in the preprocessing section.

The following attacks were used in the dataset:
\begin{itemize}
\item Bank of America attack on 09/19/2012.
\item Wells Fargo Bank attack on 09/19/2012.
\item Wells Fargo Bank attack on 09/25/2012.
\item PNC Bank attack on 09/19/2012.
\item PNC Bank attack on 09/26/2012.
\end{itemize}

\subsection{The Attack Topics}
Only the tweets from the Bank of America attack on 09/19/2012 were used in this experiment. The tweets before the attack day and on the attack day were used to train the two LDA models mentioned in the approach section.

The top, bottom 4 attack topics and their top 10 words are shown in table 1 and 2.
\begin{table}[h!]
\centering
\begin{tabular} {l|l|l}
	\hline
    Topic & Top 10 Words & SKL \\
    \hline 
      & bank, america's, prolonged, &  \\
    1 & site, slowdown, stuck, website, &9.729 \\
      & tech, slow, entirely & \\
    \hline
      & hackers, bank, nyse, &  \\
    2 & america, angered, target, sacrilegious, &9.205 \\
      & website, movie,, . & \\ 
    \hline
      & america's, bank, katherine, &  \\
    3 & fannie, mae, mangu-ward, examines, &9.099\\
      & contract, reason's, account & \\ 
    \hline
      & bank, site, america's, outage,&  \\
    4 & prolonged, sept, users, &9.055\\
      & 18, said, reported & \\  
    \hline
\end{tabular}
\caption{Top 4 Attack topics from the Bank of America data with their Symmetric Kullback-Leibler divergence}
\end{table}

\begin{table}[h!]
\centering
\begin{tabular} {l|l|l}
	\hline
    Topic & Top 10 Words & SKL \\
    \hline 
      & bank, america, follows, &  \\
    1 & light, central, sales, policy, &4.15656803208709 \\
      & check, rt, cashed & \\
    \hline
      & america, bank, bad, &  \\
    2 & great, claiming, keep, work, &4.16785261141118 \\
      & post, can, feedback & \\ 
    \hline
      & bank, america, capital, &  \\
    3 & ..., @abc, deon, pitsor, &4.30044067526549\\
      & names, >, annual & \\ 
    \hline
      & can, work, america, bank,&  \\
    4 & help, happened, anything, &4.33914718404024\\
      & jh, ma, s17 & \\  
    \hline
\end{tabular}
\caption{Bottom 4 Attack topics from the Bank of America data with their Symmetric Kullback-Leibler divergence}
\end{table}

As shown in table 1, there are roughly 4 kinds of words in the attack topics. First is the name of the entity we are watching. In this case, it is Bank of America. Those words are in every tweet, so they get very high weight in the topics, while not providing useful information. Those words can be safely discarded or added to the stop word list. The second type of words are general cybersecurity words like website, outage, hackers, slowdown and so on. Those words have the potential to become an indicator. When topics with those words appears, it is likely that there exists an attack. The third kind are words related to the specific attack but not attacks in general. Those words can provide details about the attack, but it is hard to identify them without reading the full tweets. In our example, the words movie and sacrilegious are in this group. That is because the DDoS attack on Bank of America was in response to the release of a controversial sacrilegious film. The remaining words are non-related words. The higher the weights of them in a topic, the less likely the topic is actually about a DDoS attack.

The results showed that except the 3rd topic, the top 4 topics have high weight on related words and the number of the forth type of words are smaller than the first three types of words. There are no high weight words related to security in the bottom 4 topics. We can say that the high SKL topics are about cyber attacks.

\subsection{The Attack Tweets}
In this subsection we discuss the experiment on the attack tweets found in the whole dataset. As stated in section 3.3, the whole dataset was divided into two parts. $D_a$ contained all of the tweets collected on the attack day of the five attacks mentioned in section 4.2. And $D_b$ contained all of the tweets collected before the five attacks. There are 1180 tweets in $D_a$ and 7979 tweets in $D_b$. The tweets on the attack days ($D_a$) are manually annotated and only 50 percent of those tweets are actually about a DDoS attack.

\begin{figure}
\includegraphics[width=0.5\textwidth]{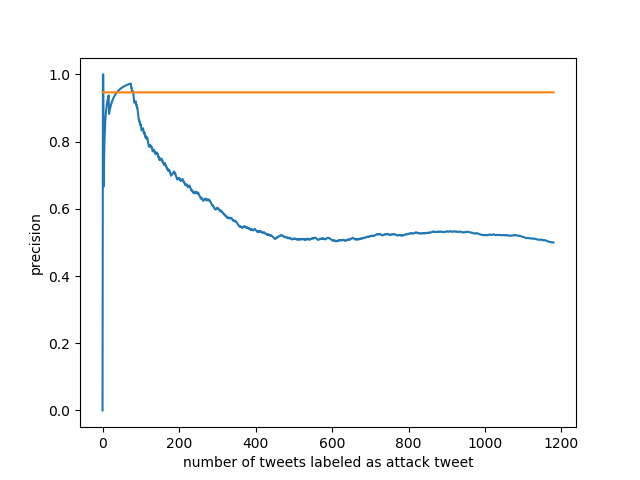}
\caption{Precision, positive predictive value, of the model when labeling the first x ranked tweets as attack tweet using all of the tweets collected. The straight line is the result of a supervised LDA model as a baseline. }
\centering
\label{precisionall}
\end{figure}

The 5 tweets that have the highest relative score in the dataset are:
\begin{itemize}
\item jiwa mines and miner u.s. bancorp, pnc latest bank websites to face access issues: (reuters) - some u.s. bancorp... http://bit.ly/p5xpmz 
\item u.s. bancorp, pnc latest bank websites to face access issues: (reuters) - some u.s. bancorp and pnc financial...
\item @pncvwallet nothing pnc sucks fat d ur lucky there's 3 pnc's around me or your bitchassness wouldnt have my money
\item business us bancorp, pnc latest bank websites to face access issues - reuters  news
\item forex business u.s. bancorp, pnc latest bank websites to face access issues http://dlvr.it/2d9ths 
\end{itemize}
The precision when labeling the first x ranked tweets as attack tweet is shown in the figure \ref{precisionall}. The x-axis is the number of ranked tweets treated as attack tweets. And the y-axis is the corresponding precision. The straight line in figures \ref{precisionall}, \ref{precisioncross} and \ref{precisionhdp} is the result of a supervised LDA algorithm which is used as a baseline. Supervised LDA achieved 96.44 percent precision with 10 fold cross validation.

The result shows that if the model is set to be more cautious about labeling a tweet as an attack tweet, a small x value, higher precision, even comparable to supervised model can be achieved. However as the x value increases the precision drops eventually.

\begin{figure}[h!]
\includegraphics[width=0.5\textwidth]{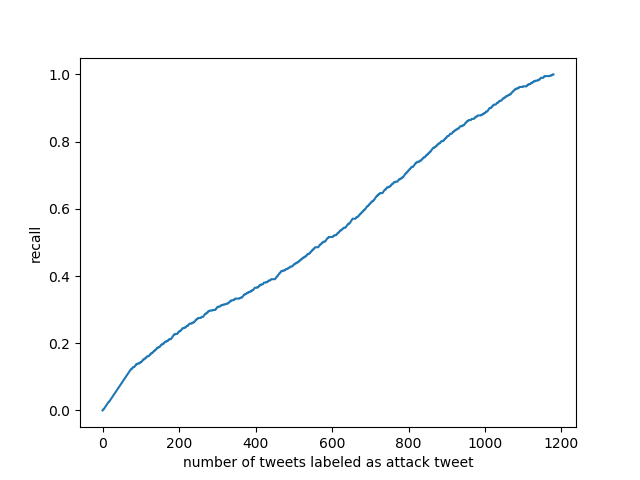}
\caption{Recall, true positive rate, of the model when labeling the first x ranked tweets as attack tweet using all of the tweets collected.}
\centering
\label{recallall}
\end{figure}

Figure \ref{recallall} shows the recall of the same setting. We can find out that the recall increases as the model becomes more bold, at the expense of precision.

\begin{figure}[h!]
\includegraphics[width=0.5\textwidth]{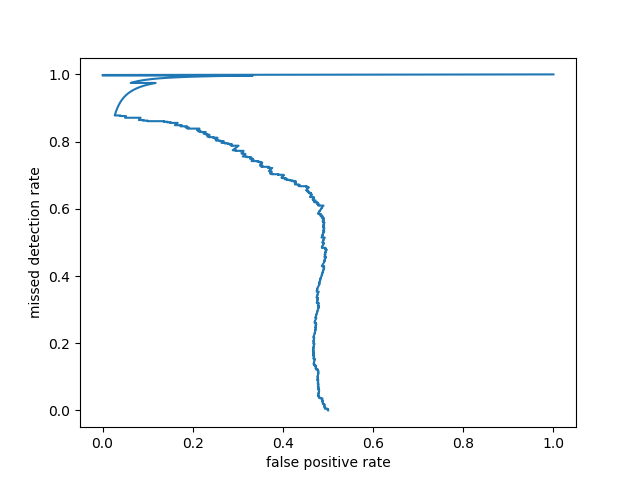}
\caption{Detection error trade-off graph when labeling the different number of ranked tweets as attack tweet using all of the tweets collected.}
\centering
\label{detall}
\end{figure}

Figure \ref{detall} is the detection error trade-off graph to show the relation between precision and recall more clearly (missed detection rate is the precision). 

\subsection{Generalization}
In this subsection we evaluate how good the model generalizes. To achieve that, the dataset is divided into two groups, one is about the attacks on Bank of America and the other group is about PNC and Wells Fargo. The only difference between this experiment and the experiment in section 4.4 is the dataset. In this experiment setting $D_a$ contains only the tweets collected on the days of attack on PNC and Wells Fargo. $D_b$ only contains the tweets collected before the Bank of America attack. There are 590 tweets in $D_a$ and 5229 tweets in $D_b$. In this experiment, we want to find out whether a model trained on Bank of America data can make good classification on PNC and Wells Fargo data.

\begin{figure}[h!]
\includegraphics[width=0.5\textwidth]{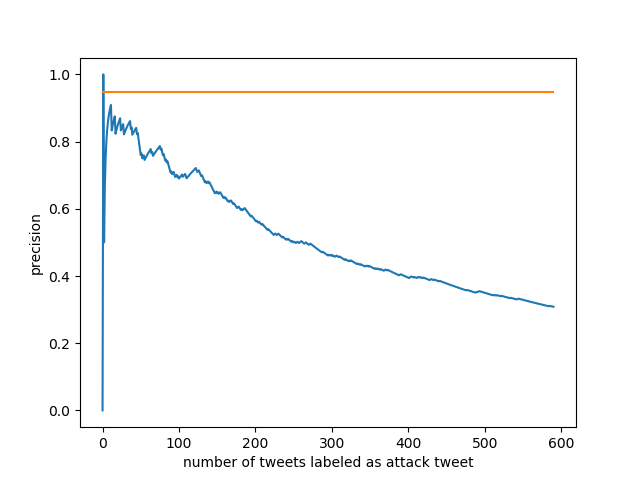}
\caption{Precision, positive predictive value, of the model when labeling the first x ranked tweets as attack tweet. The model was trained on Bank of America data and tested on PNC and Wells Fargo data. The straight line is the result of a supervised LDA model as a baseline.}
\label{precisioncross}
\centering
\end{figure}

\begin{figure}[h!]
\includegraphics[width=0.5\textwidth]{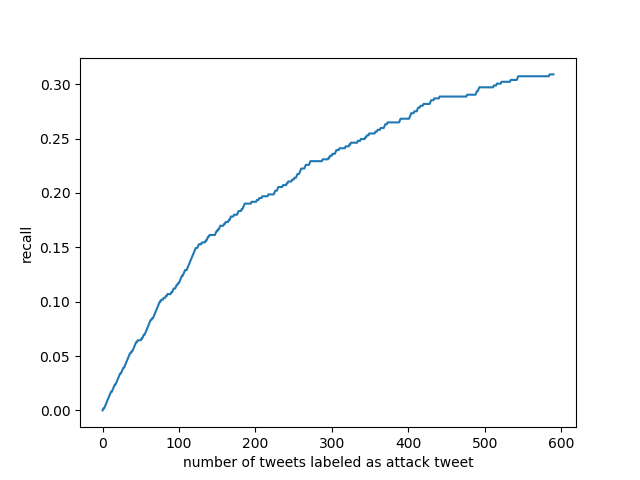}
\caption{Recall, true positive rate, of the model when labeling the first x ranked tweets as attack tweet. The model was trained on Bank of America data and tested on PNC and Wells Fargo data.}
\centering
\label{recallcross}
\end{figure}

\begin{figure}[h!]
\includegraphics[width=0.5\textwidth]{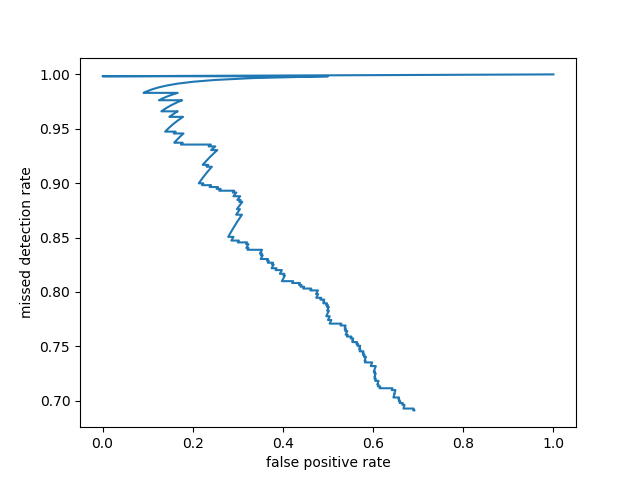}
\caption{Detection error trade-off graph when labeling the different number of ranked tweets as attack tweet. The model was trained on Bank of America data and tested on PNC and Wells Fargo data.}
\centering
\label{detcross}
\end{figure}

Figures \ref{precisioncross} and \ref{recallcross} will show the precision and recall of the model in this experiment setting. A detection error trade-off graph (Figure \ref{detcross}) is also provided.
The result is similar to the whole dataset setting from the previous section. The smaller the x value is, the higher the precision and lower the recall, vice versa. The precision is also comparable to the supervised model when a small x is chosen. This shows that the model generalized well.

\subsection{Impact Estimation}

Using the result from last section, we choose to label the first 40 tweets as attack tweets. The number 40 can be decided by either the number of tweets labeled as attack tweets by the decision tree classifier or the number of tweets that have a relative score S higher than a threshold. The PNC and Wells Fargo bank have 308.3k followers combined as of July 2018. According to eqution (5) from section 3.6, the severity Level can be computed.
\begin{equation}
SeverityLevel = \beta * {40 \over 590} + (1 - \beta) * {40 \over 308300}
\end{equation}

The score would have a range from 6.78 * $10^{-2}$ to 1.30 * $10^{-3}$, depending on the value of $\beta$. This means that it could be a fairly important event because more than six percent of tweets mentioning the banks are talking about the DDoS attack. However it could also be a minor attack because only a tiny portion of the people following those banks are complaining about the outage. The value of $\beta$ should depend on the provider's own definition of severity.

\subsection{Parameter Tuning}
This model has two parameters that need to be provided. One is $\alpha$ which is needed to determine the number of topics parameter $N_t$, and the other is whether to use the optional decision tree filter.

\begin{figure*}
\includegraphics[width=\textwidth]{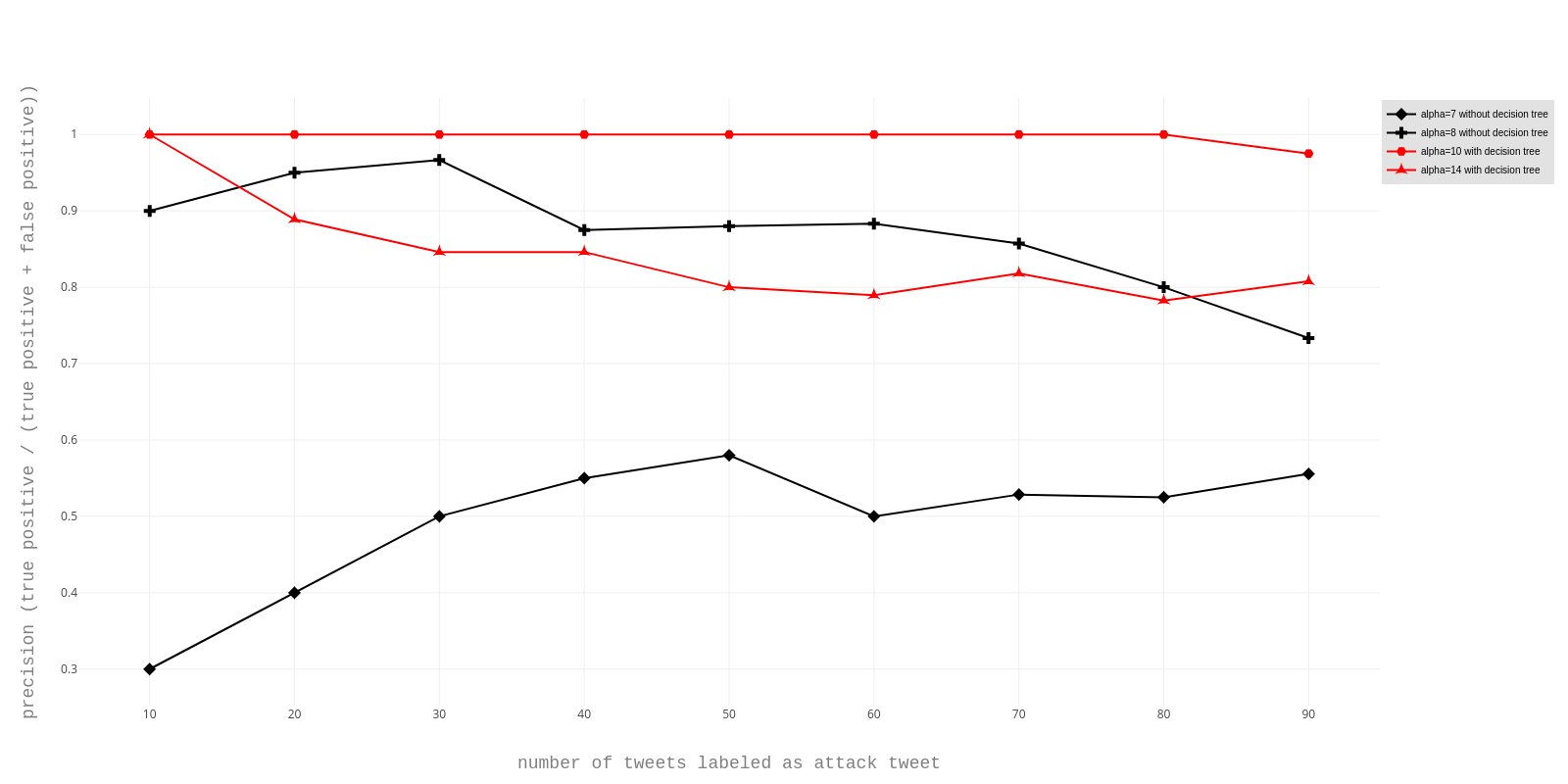}
\caption{Selected precision, positive predictive value, of the models with different parameter combinations. $\alpha$ is a parameter used to find out number of topics in the corpus. The model was trained on Bank of America data and tested on PNC and Wells Fargo data.}
\centering
\label{parameterpre}
\end{figure*}

\begin{figure*}
\includegraphics[width=\textwidth]{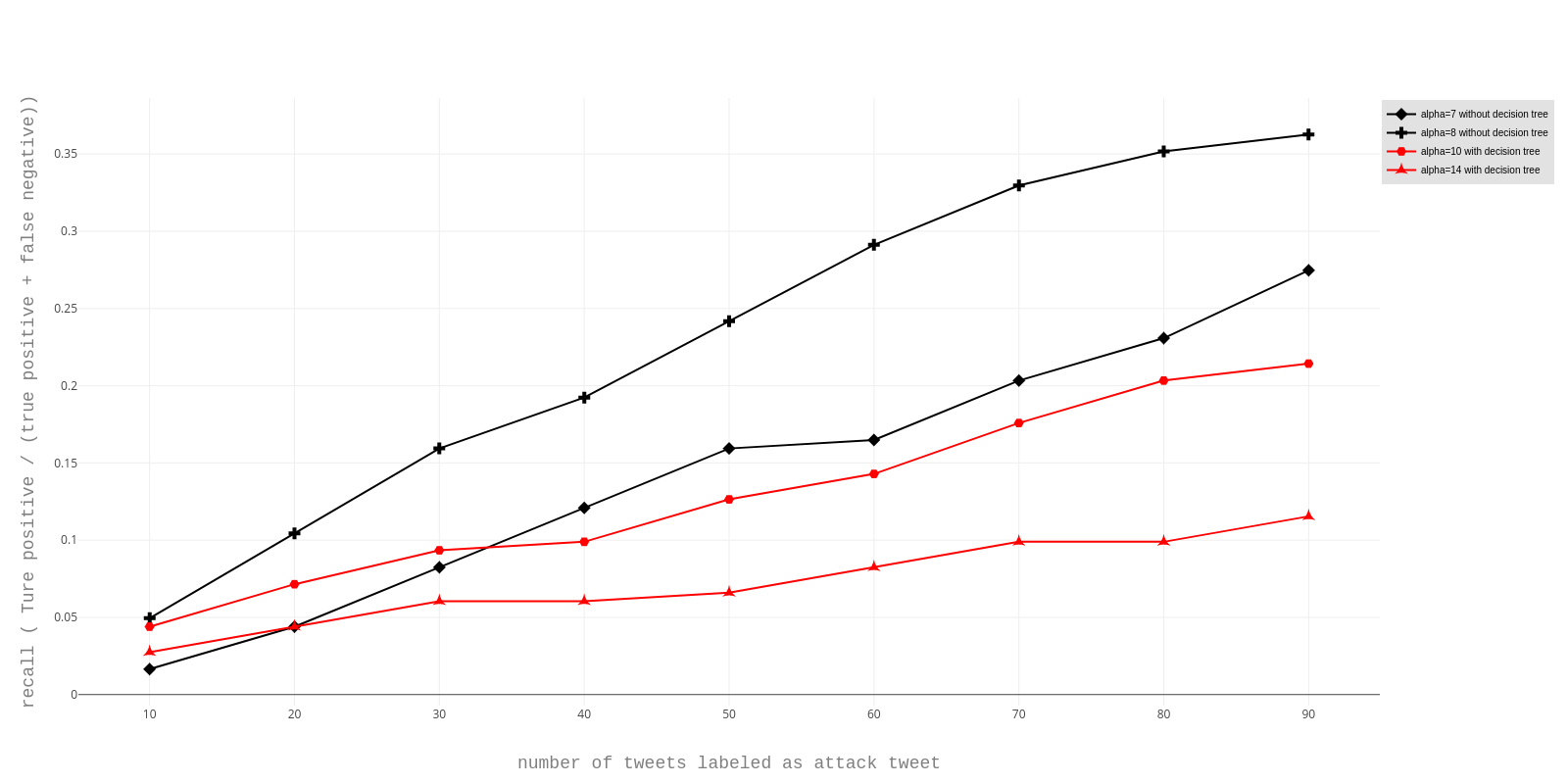}
\caption{Selected recall, true positive rate, of the models with different parameter combinations. $\alpha$ is a parameter used to find out number of topics in the corpus.  The model was trained on Bank of America data and tested on PNC and Wells Fargo data.}
\label{parameterrec}
\centering
\end{figure*}

Figures \ref{parameterpre} and \ref{parameterrec} provide experimental results on the model with different combinations of parameters. We selected four combinations that have the best and worst performance. All of the results can be found in appendix. The model was trained on Bank of America tweets and tested on PNC and Wells Fargo tweets like in section 4.5. In the figure, different lines have different values of $\alpha$ which ranges from 5 to 14 and the x axis is the number of ranked tweets labeled as attack tweets which have a range of 1 to 100 and the y-axis is the precision or recall of the algorithm and should be a number from 0 to 1.


The results shows the decision tree layer increases precision at the cost of recall. The model's performance differs greatly with different $\alpha$ values while there lacks a good way to find the optimal one.

\section{Discussion}
In this section, we will discuss two questions.

Firstly, we want to briefly discuss how good humans do on this task. What we find out is though humans perform well on most of the tweets, some tweets have proven to be challenging without additional information.  In this experiment, we asked 18 members of our lab to classify 34 tweets picked from human annotated ones. There are only two tweets which all the 18 answers agree with each other. And there are two tweets that got exactly the same number of votes on both sides. The two tweets are "if these shoes get sold out before i can purchase them, i'ma be so mad that i might just switch banks! @bankofamerica fix yourself!" and "nothing's for sure, but if i were a pnc accountholder, i'd get my online banking business done today: http://lat.ms/uv3qlo". 

\begin{figure}[h!]
\includegraphics[width=0.5\textwidth]{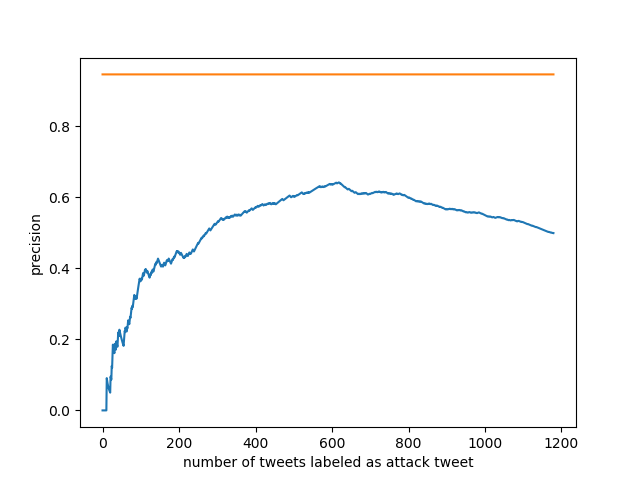}
\caption{Precision, positive predictive value, of the Hierarchical Dirichlet process model when labeling the first x ranked tweets as attack tweet using all of the tweets collected. The straight line is the result of a supervised LDA model as a baseline. }
\centering
\label{precisionhdp}
\end{figure}

\begin{figure}[h!]
\includegraphics[width=0.5\textwidth]{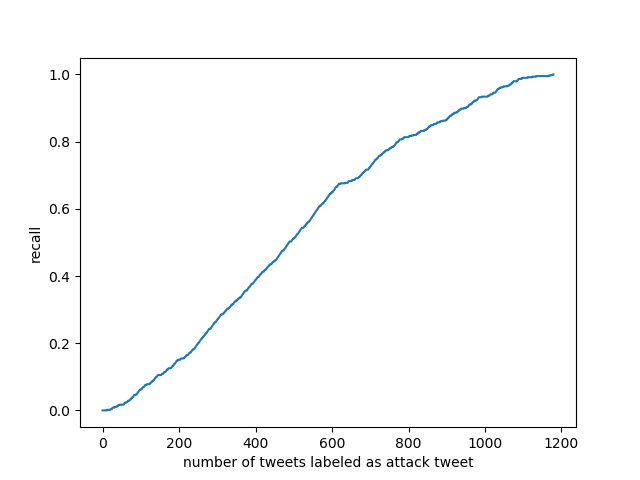}
\caption{Recall, true positive rate, of the Hierarchical Dirichlet process model when labeling the first x ranked tweets as attack tweet using all of the tweets collected.}
\centering
\label{recallhdp}
\end{figure}

The second question we want to talk about is how to find out the optimal number of topics in each of the two LDA models. As shown in the parameter tuning section, the number of topics parameter greatly affects the performance of the model. We've tried several ways to figure out the number of topics. First a set number of topics for different corpora. We tried 30 different topic numbers on the Bank of America dataset and chose the best one, and then tested it on the PNC data. The result shows that this method does not perform well on different datasets. We think it is because the number of topics should be a function of the number of documents or number of words in the corpus. Then we tried to let the model itself determines the parameter. There are some LDA variations that can do automatic number of topic inference. The one we chose is the Hierarchical Dirichlet Process (HDP) mixture model, which is a nonparametric Bayesian approach to clustering grouped data and a natural nonparametric generalization of Latent Dirichlet Allocation \cite{teh2005sharing}. However it does not perform very well. Its precision is shown in figure \ref{precisionhdp} and recall is shown in figure \ref{recallhdp}. 
We think the reason for this kind of performance might be that tweets, with the restriction of 140 characters, have very different properties than usual documents like news or articles. The last method is what was proposed in this paper. An $\alpha$ equals 10 is what we chose and did a good job on the experiments. But it is only an empirical result.

\section{Conclusion}
In this paper, we proposed a novel weakly-supervised model with optional supervised classifier layer to determine the impact of a Denial-of-Service attack in real time using twitter. The approach computes an anomaly score based on the distribution of new topics and their KL divergence to the historical topics. Then we tested the model on same and different entities to check the model's performance and how well it generalize.  Our experiment result showed that the model achieved decent result on finding out tweets related to a DDoS attack even comparable to a supervised model baseline. And it could generalize to different entities within the same domain. Using the attack tweets, we could get an estimation of the impact of the attack with a proposed formula.

There remain some interesting open questions for future research. For example, it is important to figure out a way to find out the optimal number of topics in the dataset. We would also be interested to see how well this model will perform on other kind of event detection task if the optional classifier layer changes accordingly.

\appendix
\section{Additional Result for Parameter Tuning}
\begin{figure*}
\includegraphics[width=\textwidth]{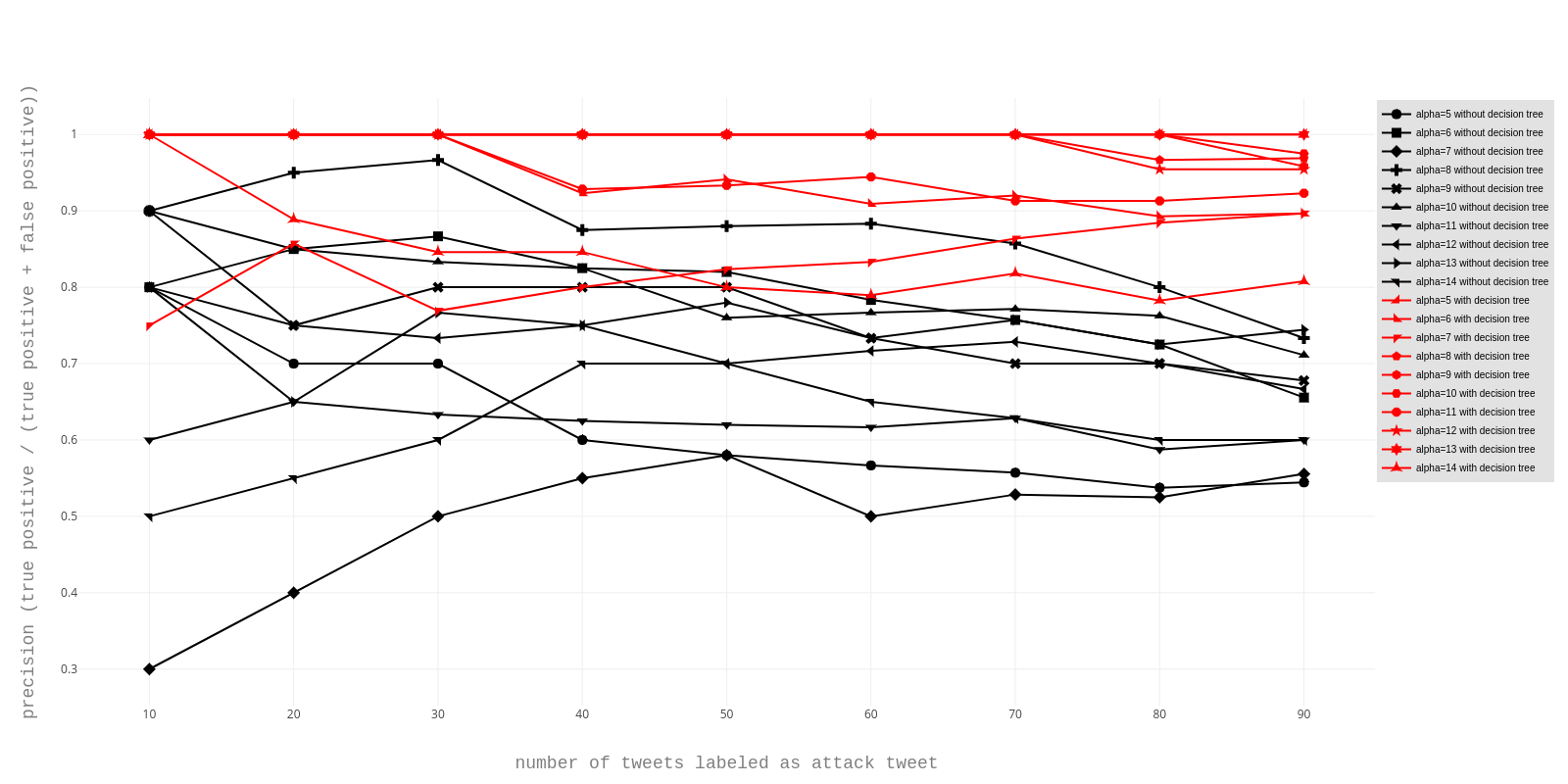}
\caption{Precision, positive predictive value, of the models with different parameter combinations. $\alpha$ is a parameter used to find out number of topics in the corpus. The model was trained on Bank of America data and tested on PNC and Wells Fargo data.}
\centering
\label{parameterpreall}
\end{figure*}

\begin{figure*}
\includegraphics[width=\textwidth]{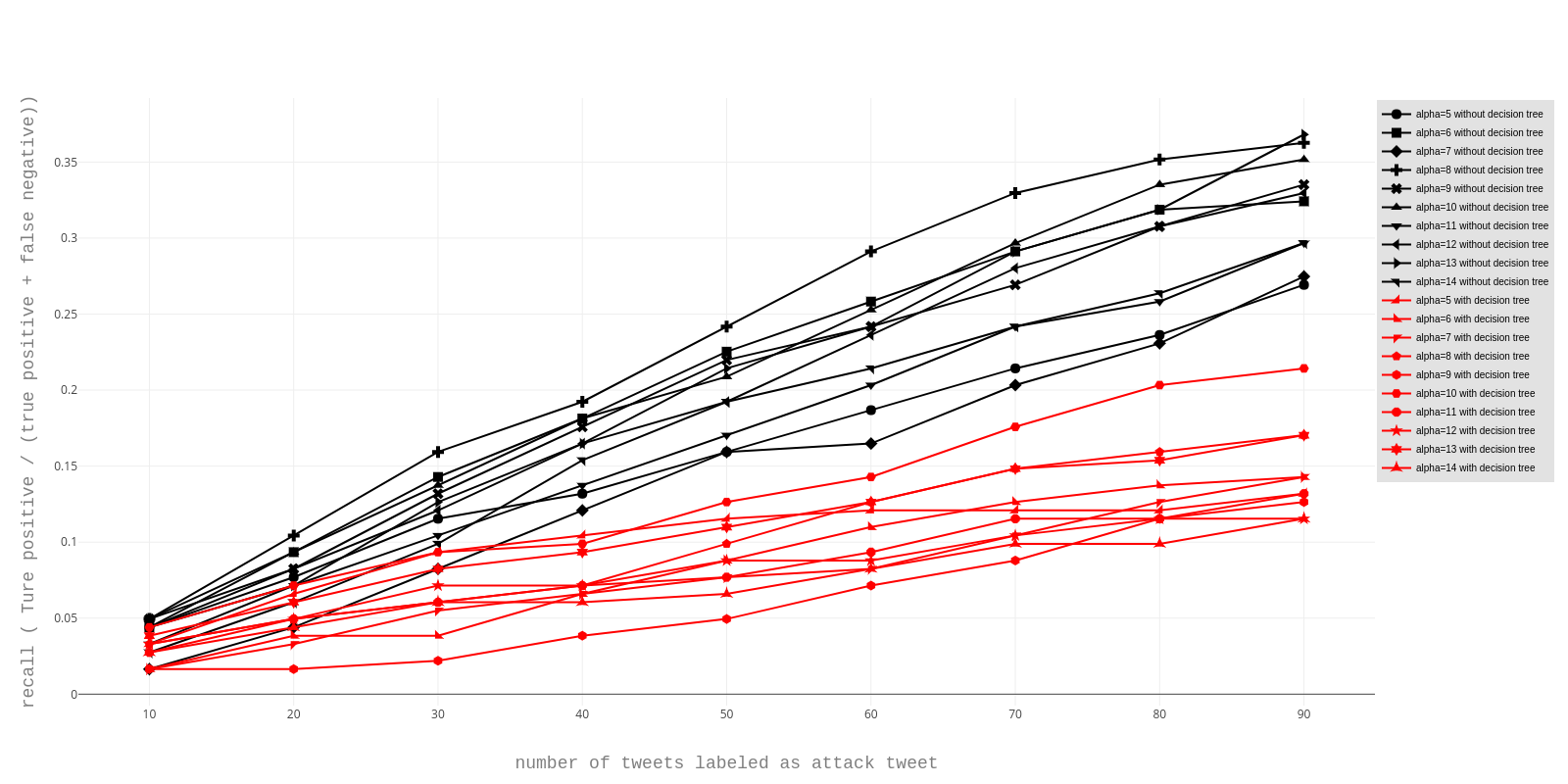}
\caption{Recall, true positive rate, of the models with different parameter combinations. $\alpha$ is a parameter used to find out number of topics in the corpus.  The model was trained on Bank of America data and tested on PNC and Wells Fargo data.}
\label{parameterrecall}
\centering
\end{figure*}

Figures \ref{parameterpreall} and \ref{parameterrecall} provide all of the experimental results on the model with different combinations of parameters. 

\bibliographystyle{ACM-Reference-Format}
\bibliography{twitter}
\end{document}